\documentclass{aa} 

\usepackage[authoryear]{natbib}
\bibpunct{(}{)}{;}{a}{}{,} % to follow the A&A style
\authorrunning{Orozco Su\'arez et al.\/} 
\titlerunning{Milne-Eddington inversions of high resolution observations of 
the quiet Sun} 

\usepackage{graphicx}
\usepackage{txfonts}
\usepackage{color}

\newcommand{\degree}{\ensuremath{^\circ}\/}
\def\farcs{\hbox{$.\!\!^{\prime\prime}$}}

\newcommand{\ms}{${\rm m}\, {\rm s}^{-1}$}
\newcommand{\kms}{${\rm km}\, {\rm s}^{-1}$}

\begin{document}
   \title{Applicability of Milne-Eddington inversions to high spatial
   resolution observations of the quiet Sun}

   \author{D.\ Orozco Su\'arez \inst{1,2}, L.~R.\ Bellot Rubio \inst{1}, 
   A.\ V\"ogler\inst{3}, \and J.~C. del Toro Iniesta\inst{1}}

   \institute{Instituto de Astrof\'{\i}sica de Andaluc\'{\i}a (CSIC),
Apdo.\ Correos 3004, 18080 Granada, Spain
\and
  National Astronomical Observatory of Japan,
2-21-1 Osawa, Mitaka, Tokyo 181-8588, Japan 
 \email{d.orozco@nao.ac.jp} 
\and  Sterrenkundig Instituut, Utrecht University, Postbus 80000, 3508 TA Utrecht, 
The Netherlands}

   \date{Received ; accepted }

\abstract{The physical conditions of the solar photosphere change on
very small spatial scales both horizontally and vertically. Such a
complexity may pose a serious obstacle to the accurate determination
of solar magnetic fields.}  {We examine the applicability of
Milne-Eddington (ME) inversions to high spatial resolution
observations of the quiet Sun. Our aim is to understand the connection
between the ME inferences and the actual stratifications of the
atmospheric parameters.}  {We use magnetoconvection simulations of the
solar surface to synthesize asymmetric Stokes profiles such as those 
observed in the quiet Sun. We then invert the profiles with the ME
approximation. We perform an empirical analysis of the heights of
formation of ME measurements and analyze the uncertainties brought about by the ME
approximation. We also investigate the quality of the fits and 
their relationship with the model stratifications.}
{The atmospheric parameters derived from ME inversions of high-spatial
resolution profiles are reasonably accurate and can be used for
statistical analyses of solar magnetic fields, even if the fit is not
always good. We also show that the ME inferences cannot be assigned to
a specific atmospheric layer: different parameters sample different
ranges of optical depths, and even the same parameter may trace
different layers depending on the physical conditions of the
atmosphere. Despite this variability, ME inversions tend to probe
deeper layers in granules as compared with intergranular lanes. }  {}


 \maketitle

%
%________________________________________________________________

 \section{Introduction} 
  \label{cap5:sec:intro}

The solar spectrum carries information about the properties of our
star. In general, a broad range of atmospheric layers contribute to
the shape of the spectral lines, making it difficult to extract this
information directly. Both the measurement process and the method of
analysis introduce uncertainties in the physical quantities retrieved
from the observations. Sources of error are photon noise and
instrumental effects like limited spectral resolution, wavelength
sampling, and angular resolution, but also the simplifications and
approximations of the model used to interpret the measurements.

In this paper we want to evaluate the merits of Milne-Eddington (ME)
inversions for the analysis of the polarization line profiles emerging
from the solar atmosphere. The ME approximation does not account for
vertical variations of the parameters \citep{1956PASJ....8..108U,
Rakk1,Rakk2}, so it cannot accurately describe the solar plasma
when rapid changes in height are present.  What is, then, the
significance of the ME parameters?

To answer this question it is necessary to simulate the processes of
line formation and data inversion. Usually one prescribes a set of
model atmospheres, performs spectral synthesis calculations, inverts
the synthetic profiles, and compares the results with the known input.
A common approach is to use ME models both to generate the spectra and
to invert them \citep[e.g.,][]{nortonetal2006, borreroetal2007}. In
that case the analysis is internally consistent and the uncertainties
of the retrieved ME parameters are mostly due to the noise and, to a
smaller extent, to the convergence of the algorithm, provided
that the spectral resolution and wavelength sampling are
appropriate. Uncertainties caused by photon noise are known as
statistical errors and can be evaluated by means of numerical tests
or, more efficiently, by using ME response functions
\citep{2007A&A...462.1137O,2010ApJ...711..312D}. However, they represent 
only a small fraction of the total error. Another source of error is
the very assumption of height-independent parameters, which leads
to symmetric line profiles. What happens when realistic (i.e.,
asymmetric) Stokes spectra are analyzed in terms of ME models? Do the
uncertainties of the retrieved parameters increase significantly?
Answering these questions is the aim of the present work.

\begin{figure*}[!t] \centering 
\resizebox{0.425\hsize}{!}{\includegraphics{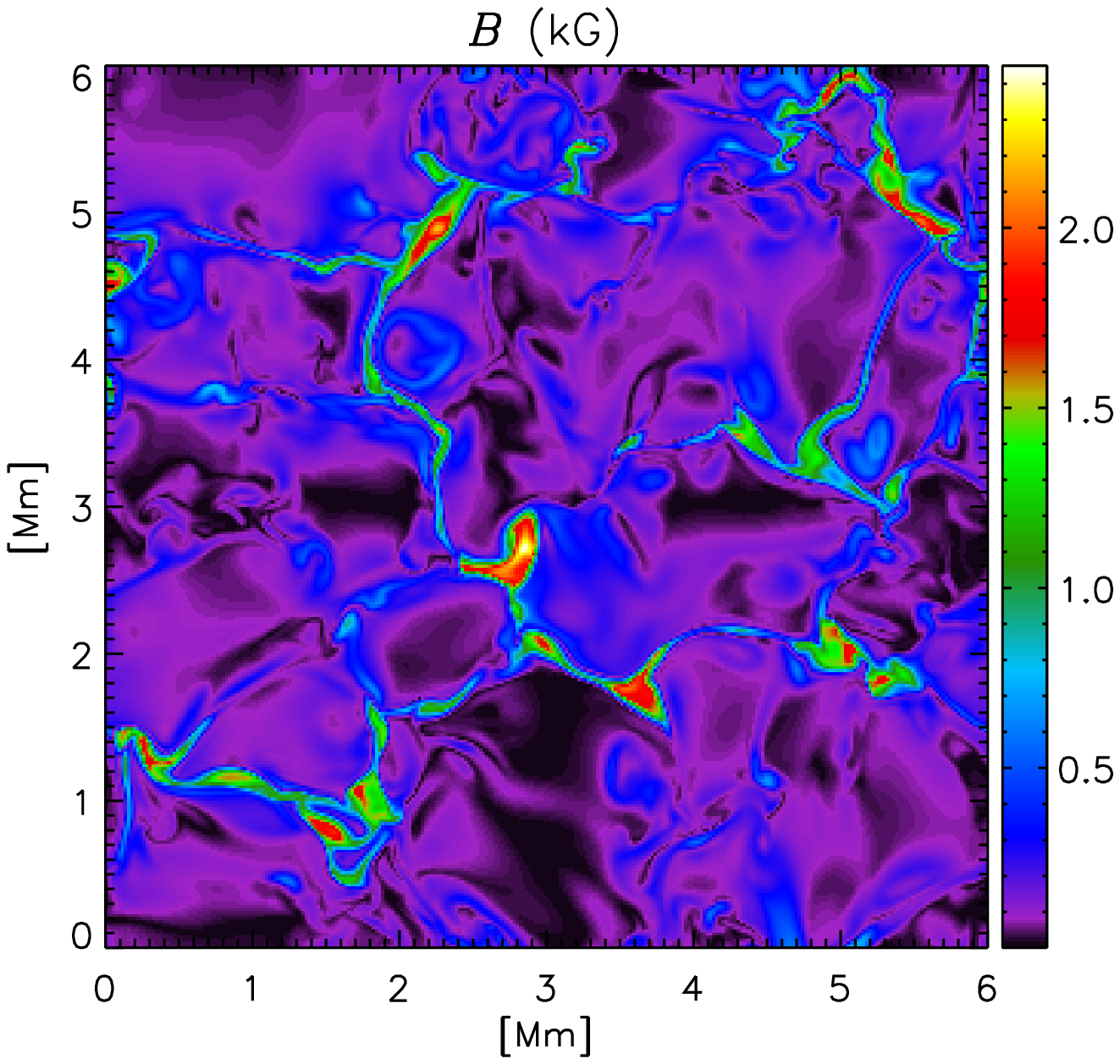}}
\resizebox{0.425\hsize}{!}{\includegraphics{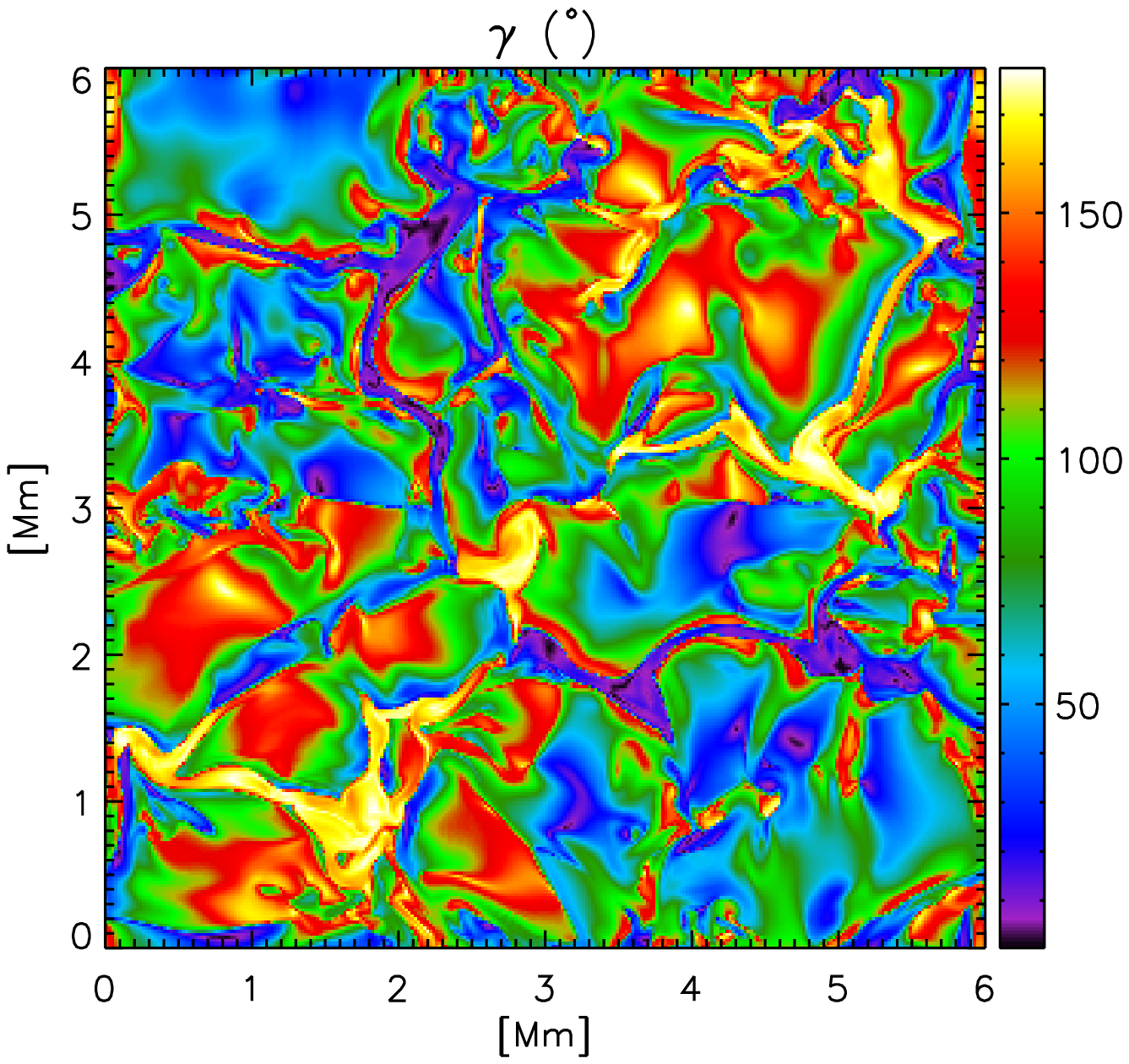}}
\resizebox{0.425\hsize}{!}{\includegraphics{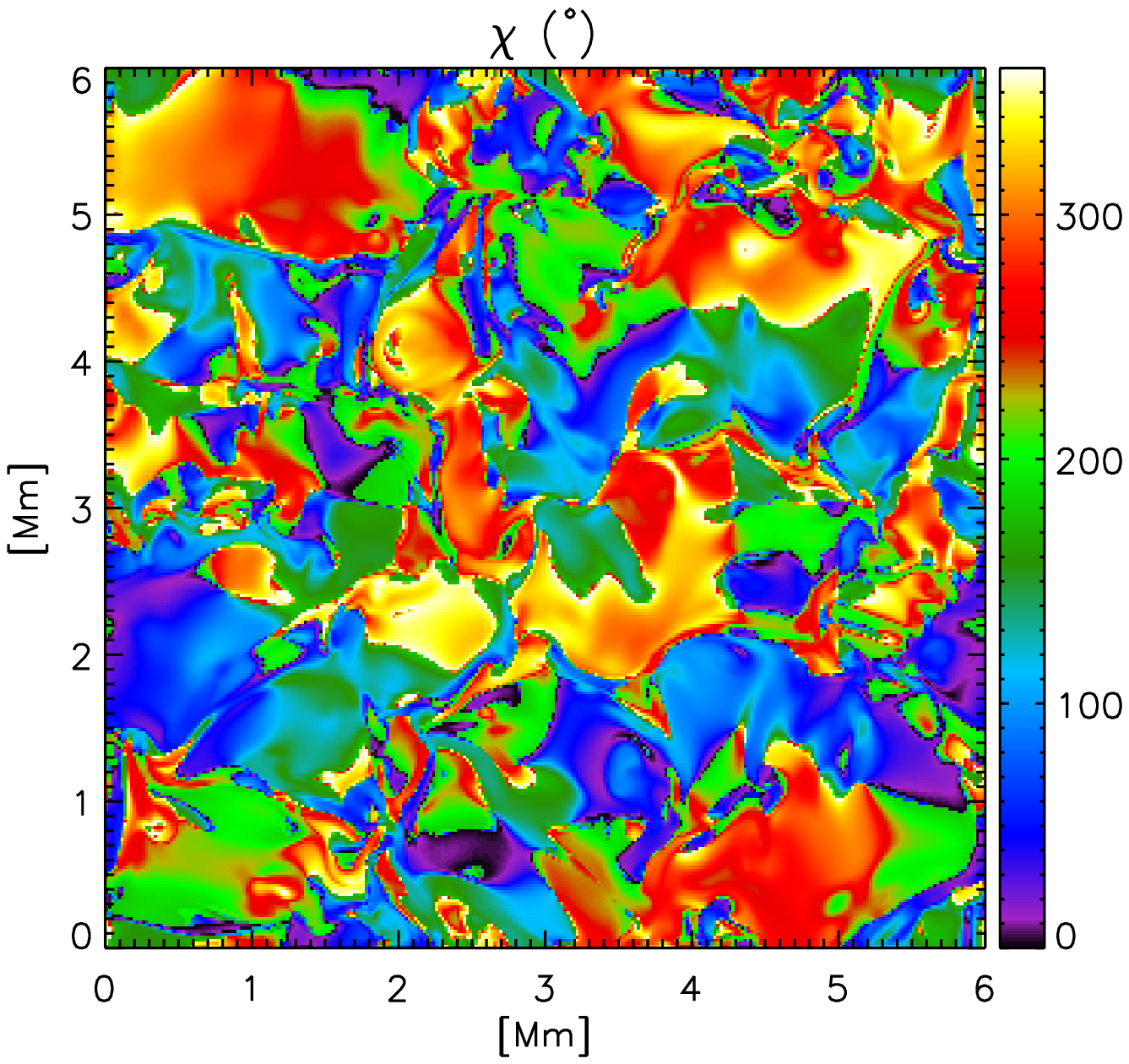}}
\resizebox{0.425\hsize}{!}{\includegraphics{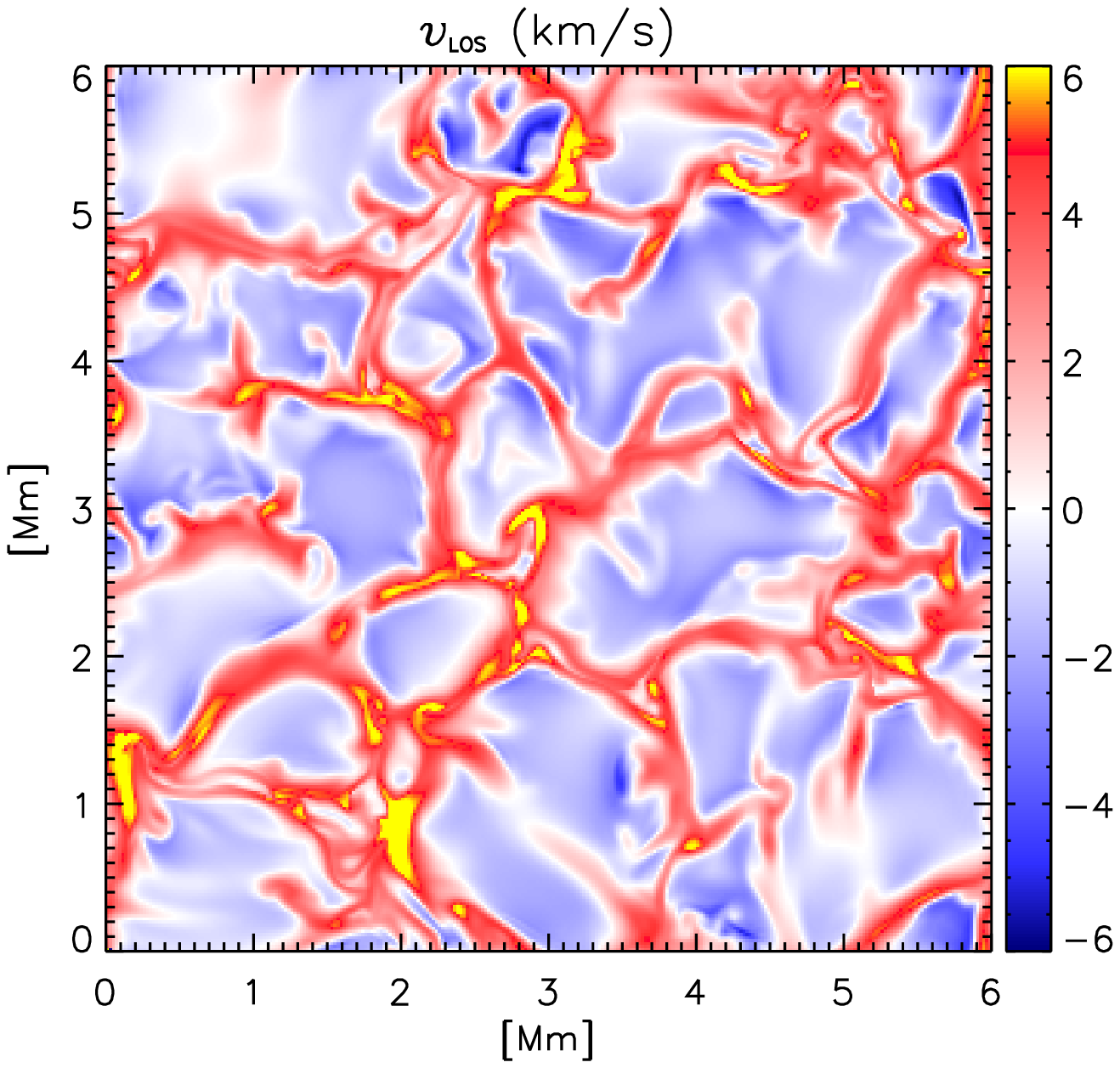}}
\caption{Magnetic field strength, inclination, azimuth, and 
LOS velocity at $\tau=1$ in a simulation snapshot with $\langle B
\rangle=140$~G. Negative velocities represent upflows.}
\label{Fig1} 
\end{figure*}

A first study of the capabilities and limitations of ME inversions was
carried out by \cite{1998ApJ...494..453W} using simple (non-ME) model
atmospheres. They made a quantitative comparison of results obtained
with the ME code of the High Altitude Observatory
\citep{lites2,lites1} and the SIR code \citep[Stokes Inversion based
on Response functions;][]{1992ApJ...398..375R}. The main conclusion of
their work was that ME inversions provide accurate values of the
physical parameters averaged along the line of sight, at least when
the stratifications are smooth.

More recently, \cite{2007MmSAI..78..166K} have investigated whether
the magnetic field stratification itself can be determined reliably
through inversion of high resolution data. To that end, they
synthesized the Stokes profiles of the \ion{Fe}{i} 630~nm lines with
the help of MHD models and inverted them with SIR, allowing for
vertical gradients of the atmospheric parameters. The analysis showed
that SIR is able to recover the actual magnetic stratification for
fields as weak as 50~G if no noise is present. This work extends the
results of \cite{1998ApJ...494..453W} to the case in which the
stratifications are not smooth.

To determine the uncertainties associated with ME inversions of
asymmetric Stokes profiles we use state-of-the-art magnetohydrodynamic
simulations (Sect.~\ref{MHD}). Our goal is to describe the solar
photosphere as realistically as possible. We construct model
atmopheres from the simulations and synthesize the Stokes profiles of
the \ion{Fe}{i} 630.2~nm lines emerging from them
(Sect.~\ref{synthesis}). The SIR code is used for the spectral
synthesis, so the profiles are asymmetric. Finally, we apply a ME
inversion to the data (Sect.~\ref{inversion}). In our numerical
experiments, the spatial sampling of the MHD models, $0\farcs 0287$,
is preserved. There are two reasons why we neglect the effects of
solar instrumentation: first, they have already been studied in the
past (e.g., Orozco Su\'arez et al. 2007, 2010); second, this sampling
is close to critical for the observations to be delivered by large
telescopes like the Advanced Technology Solar
Telescope\footnote{http://atst.nso.edu/} \citep{wagner2006} and the
European Solar Telescope\footnote{http://www.iac.es/project/EST/}
\citep{collados2008}. We invert the profiles with
the MILOS code \citep{2007A&A...462.1137O}.\footnote{MILOS is
programmed in IDL and can be downloaded from our website, http://spg.iaa.es/download.asp} A direct comparison of the retrieved
and true parameters allows us to determine the effective ``heights of
formation'' of the ME parameters (Sect.~\ref{results}) and to quantify
the errors caused by the ME approximation (Sect.~\ref{results2}). The
conclusions of our work are given in Sect.~\ref{conclus}.  For
completeness, the results of ME inversions are compared with those of
tachogram/magnetogram-like analyses in the Appendix.

%------------------------------------------------------------------------

\section{Magnetohydrodynamic simulations}
\label{MHD}

We use radiative MHD simulations performed with MURaM, the
MPS/University of Chicago RAdiative MHD code
\citep{voegler, 2005A&A...429..335V}. This code solves the 3D 
time-dependent MHD equations for a compressible and partially ionized
plasma taking into account non-grey radiative energy transport and
opacity binning \citep{1982A&A...107....1N}.

Among other problems, MURaM has been employed to study facular
brightenings \citep{2004ApJ...607L..59K}, the relation between G-band
bright points and magnetic flux concentrations \citep{schussler,
2004A&A...427..335S}, the emergence of magnetic flux tubes from the
upper convection zone to the photosphere \citep{2006cheung tesis,
2007A&A...467..703C}, the strongly inclined magnetic fields of the
internetwork \citep{2008A&A...481L...5S}, umbral dots
\citep{2006ApJ...641L..73S}, solar pores \citep{2007A&A...474..261C},  
and even full sunspots \citep{2009ApJ...691..640R,
2009Sci...325..171R}. MURaM has also been used to evaluate the
diagnostic potential of spectral lines
\citep{2007ApJ...659.1726K, 2007MmSAI..78..166K, khomenko,
2005A&A...436L..27K,2007A&A...469..731S}, the validity of visible
lines for the study of internetwork magnetic fields at high spatial
resolution \citep{2007ApJ...662L..31O}, and the continuum contrast of
the solar granulation \citep{2008A&A...484L..17D}.

In this paper we consider a 5-minute sequence of a mixed-polarity
simulation run representing a network region with an average
magnetic field strength $\langle B \rangle=140$~G at $\log \,
\tau=-1$.\footnote{All optical depths refer to the continuum opacity at
500~nm} The cadence is 10~s, so we have 30 snapshots. A bipolar
distribution of vertical fields with $\langle B \rangle=200$~G was
used to initialize the simulations. Additional details about this
particular run can be found in \citet{khomenko}.

The computational box has $288\times 288 \times 100$ grid points and
covers 6000 km in the horizontal direction and 1400 km in the vertical
direction. The model extends from $z=-800$ to $z=600$~km, with
$z=0$~km the average of the heights where $\tau=1$. The spatial
grid sampling is 0\farcs0287, implying an equivalent
resolution of 0\farcs057 (41.6~km) on the solar surface. The
simulation provides the density, the linear momentum density
vector, the total energy density, the magnetic field vector, the
temperature, and the gas pressure at every grid point. The
time-averaged radiation flux density leaving the top of the box has
the solar value $F_{\odot}=6.34\times10^{10}$ erg~s$^{-1}$~cm$^{-2}$.

  \section{Spectral synthesis}
\label{synthesis}

In order to compute synthetic Stokes profiles we solve the radiative
transfer equation (RTE) for polarized light. The calculations are
carried out using the SIR code with the opacity routines of
\cite{1974SoPh...35...11W}. The spectral synthesis is accomplished in
two steps: first, the input model atmospheres are built from the MHD
simulations; then, the RTE is solved.

%------------------------------------------------------------------------

  \subsection{MHD models and spectral line synthesis}

The parameters needed for the spectral synthesis are the temperature,
electron pressure, line-of-sight (LOS) velocity, magnetic field
strength, inclination and azimuth, and optical depth. The simulations
provide most of them. However, the electron pressure and
optical depth need to be computed from the local temperature, gas
pressure, and density by solving the Saha and Boltzmann equations. The
optical depth scale is set up assuming that $z=600$~km (the top of the
computational box) corresponds to $\log\tau=-4.9$. This value has been
taken from the Harvard-Smithsonian Reference Atmosphere
\citep[HSRA;][]{Gingerich}. Finally, the resulting stratifications 
are interpolated to an evenly spaced optical depth grid using
second-order polynomials. The grid extends from $\log{\tau}=-4$ to $2$
with $\Delta\log\tau=0.05$. This range of optical depths encompasses
the formation region of all photospheric lines, except the cores
of the strongest ones.

Figure \ref{Fig1} displays maps of the field strength, inclination,
azimuth, and LOS velocity at $\tau=1$ for one simulation snapshot. In
the velocity map the granulation pattern is clearly visible, with
granular upflows that are weaker than the intergranular
downflows. Some of the small-scale intergranular structures 
have velocities of up to 6~\kms.

The field strength map shows strong flux concentrations in the
intergranular lanes. Granules also harbor magnetic fields, but they
seldom exceed 300~G.  There is a tight correlation between the
field strength and inclination in these simulations: the intergranular
fields tend to be vertical, whereas the granules exhibit more
horizontal fields. Finally, the azimuth map is dominated by 
granular-sized structures with diameters of 1\arcsec-2\arcsec\/ 
(0.7-1.5~Mm).

  \begin{figure} 
  \centering 
   \resizebox{1\hsize}{!}{\includegraphics{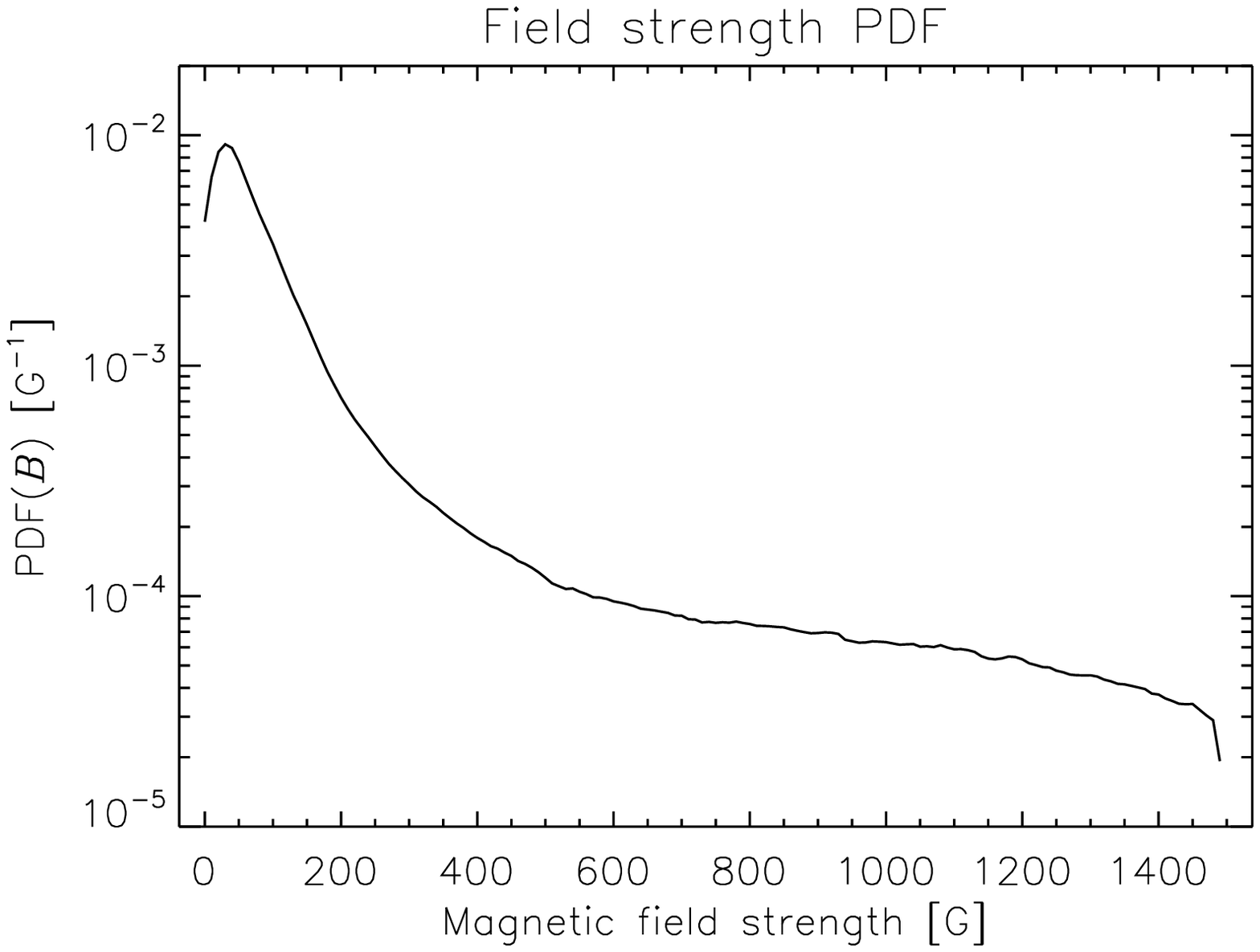}}
   \resizebox{1\hsize}{!}{\includegraphics{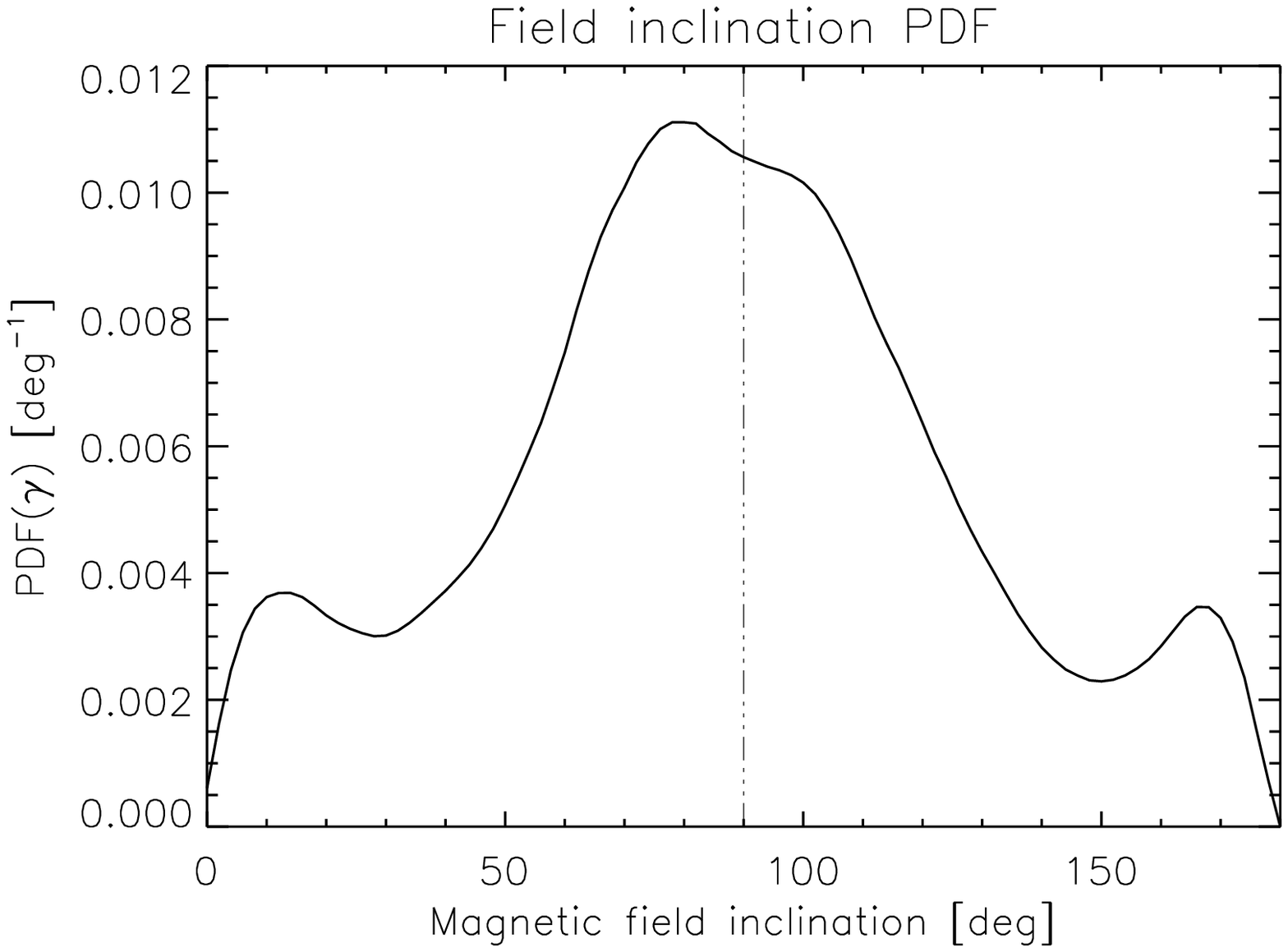}}
  \caption{Probability density functions for the magnetic field strength 
  and field inclination at $\log\tau=-1$.}
  \label{Fig2} 
  \end{figure}

  Figure~\ref{Fig2} depicts the probability density functions
  (PDFs)\footnote{The PDF is defined such that $P(B)dB$ is the
    probability of finding a magnetic field $B$ in the interval
    [$B,B+dB$]. The integral of the PDF is unity, i.e.,
    $\int_0^{\infty} P(B)dB = 1$.}  of the magnetic field strength and
  inclination at optical depth $\log\tau=-1$, averaged over the
    30 available snapshots.  The field strength PDF increases rapidly
  toward weak fields, indicating that most pixels have magnetic fields
  of the order of hectogauss. The distribution peaks at about 20~G.
  The inclination PDF shows some vertical fields and a larger
  occurrence of horizontal fields. The simulation run was seeded with
  mixed-polarity vertical fields; therefore, the distribution is
  rather symmetric about $\gamma=90\degree$.

\begin{table*}
\caption{\label{cap5:tablelines}Atomic parameters of the 
spectral lines.}
\centering
\begin{tabular}{@{}c c c c c c c c c c@{}} \hline  {\scshape Ion} & $\lambda$
   (nm) & $\chi_{\mathrm{low}}$ & $\log (gf)$ &  {\scshape Transition }& $\alpha$ & $\sigma/a_0^2$ & $g_\mathrm{eff}$\\ 
\hline \hline 
\ion{Fe}{i} & 630.1501 & 3.654 &  $-0.75$  & $5P_2-5D_2$ & 0.243 & 840.5 & 1.67 \\ 
\ion{Fe}{i} & 630.2494 & 3.686 &  $-1.236$ & $5P_1-5D_0$ & 0.240 & 856.8 & 2.5 \\ 
\hline \end{tabular}
\tablefoot{
$\lambda$ stands for the central wavelength of the
transition, $\chi_{\mathrm{low}}$ is the excitation potential of the
lower atomic level in eV, $\log gf$ represents for the multiplicity of
the lower level times the oscillator strength, $\alpha$ and $\sigma$
are the collisional broadening parameters in the quantum theory of
Anstee, Barklem and O'Mara (in units of Bohr's radius, $a_0$), and 
$g_\mathrm{eff}$ is the effective Land\'e factor of the transition.
}
\end{table*}

Once we have constructed model atmospheres for each of the $288\times
288$ pixels and for all the snapshots, we use them to compute the
Stokes profiles of \ion{Fe}{i} 630.15 and 630.25~nm. The atomic parameters used in the calculations are given in
Table~\ref{cap5:tablelines}.  The $\log gf$ values have been taken
from the VALD database \citep{Piskunov}, except for \ion{Fe}{i}
630.25~nm which is not available in VALD and comes from a fit to the
solar spectrum using the two-component model of
\cite{borrero2002}. The collisional broadening coefficients $\alpha$
and $\sigma$ due to neutral hydrogen atoms have been evaluated
following the procedure of \cite{Anstee} and
\citet{Barklem1,Barklem2}. The abundances have been taken from 
\cite{thevenin}, i.e., a value of 7.46 is employed for iron.

%------------------------------------------------------------------------

\subsection{Synthesis results}

\begin{figure}[!t] \centering 
 \resizebox{\hsize}{!}{\includegraphics{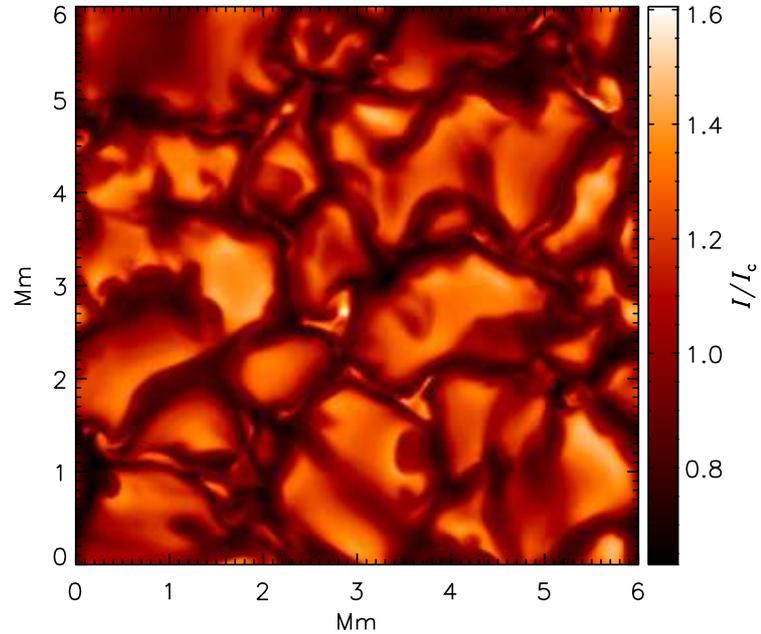}}
\caption{Continuum intensity map for the simulation snapshot 
depicted in Fig.~\ref{Fig1}. The wavelength is 630~nm.}
\label{Fig3}\end{figure}

Figure \ref{Fig3} shows a continuum map of the simulated region. The
rms contrast, computed as the standard deviation divided by the mean,
is 14.8\% at 630~nm. This contrast exceeds the
typical values obtained from ground-based observations; the
speckle-reconstructed G-band images taken at the Dunn Solar Telescope,
for example, show contrasts of 14.1\%
\citep{2007ApJ...668..586U}. Nevertheless, the agreement 
is reasonable because the observed values are degraded by instrumental
effects \citep[e.g.,][]{2008A&A...484L..17D, 2009A&A...503..225W}.

\begin{figure}[!t] \centering
  \resizebox{\hsize}{!}{\includegraphics{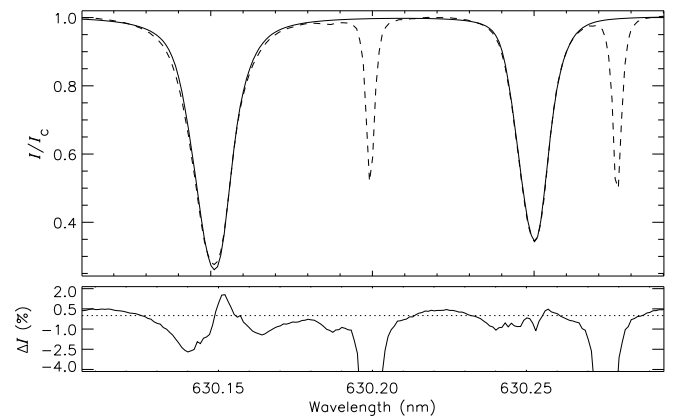}}
  \caption{Average intensity profiles from the simulations (solid) and
  the FTS atlas (dashed), for the \ion{Fe}{i} lines at 630.2~nm. The bottom panel shows the intensity differences (FTS  - simulation) in percent. }
\label{Fig4}
  \end{figure}

We have compared the temporally and spatially averaged intensity
profiles with the corresponding ones in the Fourier Transform
Spectrometer (FTS) atlas of the quiet Sun by \cite{Brault1} and
\cite{Brault2} (see Fig.~\ref{Fig4}). For the comparison, the
simulated profiles have been first normalized to unity and then
displaced in wavelength to correct for the solar gravitational
redshift \citep[see ][]{1980SoPh...66..245L}. Also, an additional
minor correction to the wavelength shift has been allowed to improve
the fits. The figure shows that the widths of the synthetic profiles
are very similar to those measured in the FTS atlas. However, the
bottom panels indicate that the simulations do not completely
reproduce the line asymmetries: the differences between observed and
synthetic profiles are small (of the order of 2.5\%), but not zero.

In summary, despite some differences between the FTS and the synthetic
profiles, the simulations appear to explain the observations rather
well. Therefore, we hope that they also produce sufficiently realistic
polarization spectra, in particular with regard to their asymmetries.

\section{ME inversion of the Stokes profiles}
\label{inversion}

To determine the vector magnetic field and the LOS velocity we
perform ME inversions. Given the high spatial and temporal
resolution of the simulations, the macroturbulence is set to zero and
the filling factor to unity, i.e., the magnetic atmosphere is assumed
to occupy the whole pixel (one-component model atmospheres).  A total
of 9 quantities are determined from the inversion: the thermodynamic
parameters $S_0$, $S_1$, $\eta_0$, $\Delta\lambda_D$, and $a$
(representing the intercept and the slope of the source function, the
line-to-continuum opacity ratio, the Doppler width, and the damping
parameter), the strength, inclination, and azimuth of the magnetic
field vector ($B$, $\gamma$, and $\chi$), and the line-of-sight velocity
($v_{\rm LOS}$). The Stokes profiles are taken from a single snapshot
of the simulation. No noise is added to the spectra to better isolate
the uncertainties due to the ME approximation.

The \ion{Fe}{i} lines at 630~nm belong to the same multiplet and
are formed under very similar thermodynamic conditions. Hence, we can
reliably assume that the ME thermodynamic parameters are the same for
both lines except for $\eta_0$. A simultaneous inversion with no extra
free parameters is thus possible using a constant $\eta_0$ ratio and
the same $\Delta\lambda_D$ and $a$ values for the two lines. This is
the strategy implemented in several ME codes
\citep[for details, see][]{Orozco_rn}.

We use the same initial guess model for all the pixels, stopping 
the inversion when convergence is achieved or 200 iterations have 
been performed. The initial model is given by $S_0=0.2$, $S_1=0.8$,
$\eta_0=6.5$, $\Delta \lambda_\mathrm{D}=30$ m{\AA}, $a=0.03$,
$B=200$~G, $\gamma=20$\degree, $\chi=20$\degree, and
$v_\mathrm{LOS}=0.25$ \kms\/.

%---------------------------
%---------------------------
%---------------------------

\section{Understanding ME inferences}
\label{results}

\begin{figure*} \centering
\resizebox{.8\hsize}{!}{\includegraphics{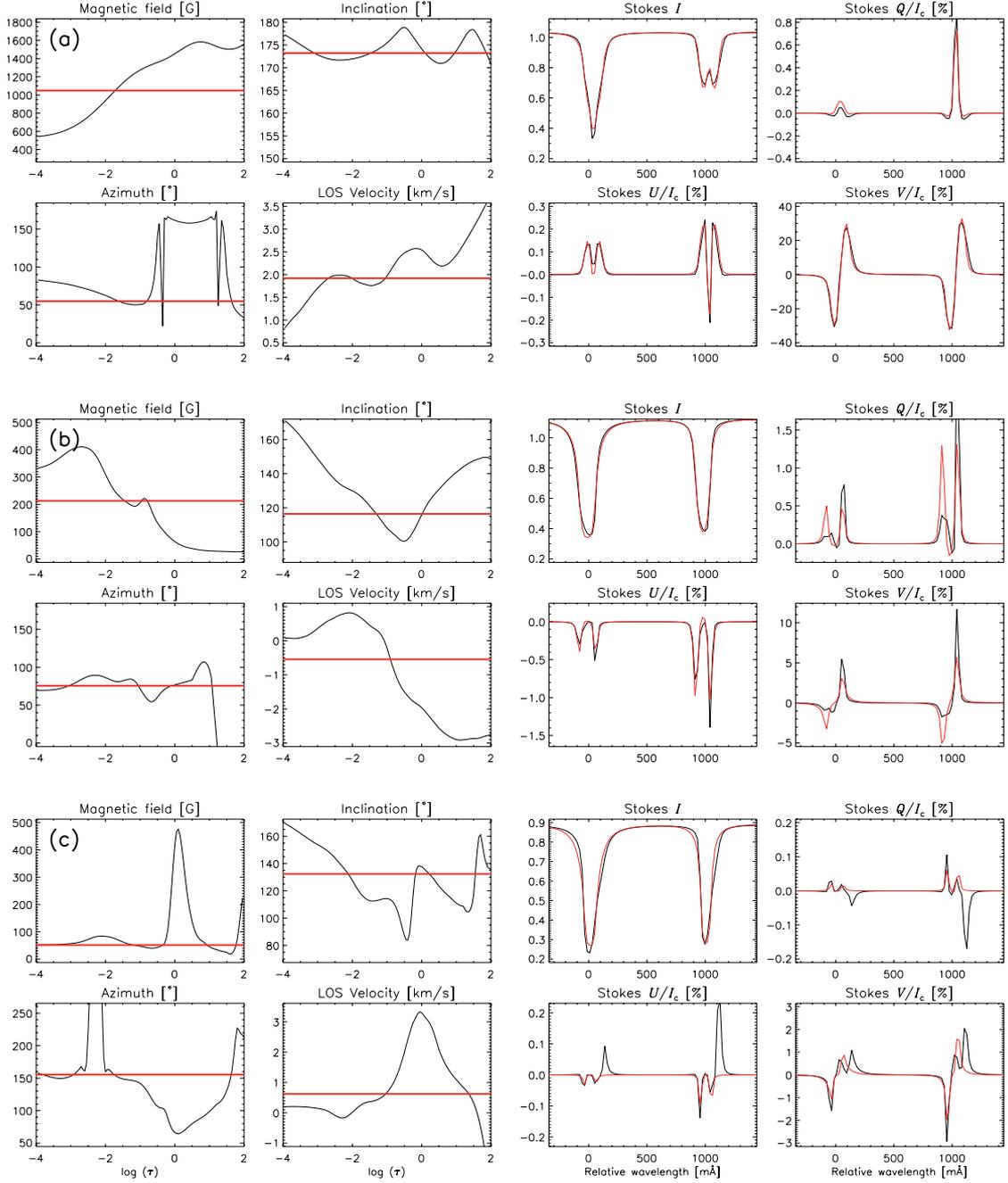}}
  \caption{Examples of MHD atmospheres and simulated profiles (black)
  and ME fits (red) for three different pixels. Left panels:
  magnetic field strength, inclination, azimuth, and LOS velocity 
  stratifications. The red horizontal lines indicate the
  inversion results. Right panels: Stokes $I$, $Q$, $U$ and $V$
  profiles synthesized from the MHD simulations with no noise (black)
  and ME fits (red). Cases (a), (b) and (c) correspond to (x,y)=(1.75,
  1.08), (2.90, 2.58), and (1.67, 1.73) Mm in Figs.~\ref{Fig6} 
  and \ref{Fig7}. }
\label{Fig5}
\end{figure*}

The line asymmetries observed in the solar atmosphere imply that there
are vertical gradients of the physical parameters, because the line
wings trace relatively deep layers and the core is formed higher in
the atmosphere. By contrast, the ME fits are strictly symmetric and
deliver height-independent atmospheric parameters. Thus, it is 
important to verify that the ME model can indeed be used for 
the analysis of observations at very high spatial resolution.

Figure \ref{Fig5} shows the magnetic field strength, inclination,
azimuth, and LOS velocity stratifications in three pixels of the
simulation snapshot, labeled (a), (b), and (c). The figure also
displays the corresponding Stokes $I$, $Q$, $U$, and $V$ profiles. The
results of the ME inversions are overplotted in red. Case (a) shows
symmetric Stokes profiles, in (b) the profiles are rather asymmetric,
and (c) shows three-lobed Stokes $V$ spectra together with anomalous
linear polarization signals. (a) represents a strong field case and
(b) and (c) correspond to weak fields. In the three examples the
atmospheric quantities undergo large variations with optical depth.

The ME fit is good in (a) and worse in (b) and (c). Clearly, as the
asymmetry level increases, the ME model has more difficulties to 
reproduce the profiles. The misfits are obvious in Stokes $Q$, $U$,
and $V$, and less dramatic in Stokes $I$.

The models retrieved from the inversion are shown in the left panels
of Fig.~\ref{Fig5} (red lines). The height-independent ME parameters
can be interpreted as averages of the actual stratifications weighted
by the corresponding response functions
\citep{1998ApJ...494..453W}. In general it is difficult to confirm 
this by simply looking at the atmospheric stratifications, but case
(c) provides a particularly clear example. This case represents a
pixel whose ME fit is not satisfactory. The analysis of the
stratifications shows that the profiles arise from an atmosphere that
has sharp discontinuities in field strength, inclination,
azimuth, and LOS velocity, located more or less at the same optical
depth. The two parts of the atmosphere separated by the discontinuity
leave clear signatures in the emergent Stokes $V$ spectra, to the
point that the magnetic and kinematic properties of the plasma
above and below the discontinuity can roughly be guessed from a simple
inspection of the profiles: there is a weak field associated with
small velocites and a stronger field showing large
redshifts. Surprisingly, however, the ME model returned by the
inversion seems to describe only the weak field.

To understand why this happens, it is important to realize that the
inversion algorithm uses all the wavelength samples to determine the
best-fit ME parameters. As mentioned before, different wavelength
positions across the line trace different atmospheric layers. Thus,
the ME inversion is forced to return average parameters along the LOS
in order to fit the whole line profile reasonably well without any
bias toward better fits in the core or the wings. This favors the weak
field component of the atmosphere because it occupies most of the
line-forming region in this particular example. 

\begin{figure*}[!t] \centering
\resizebox{0.42\hsize}{!}{\includegraphics{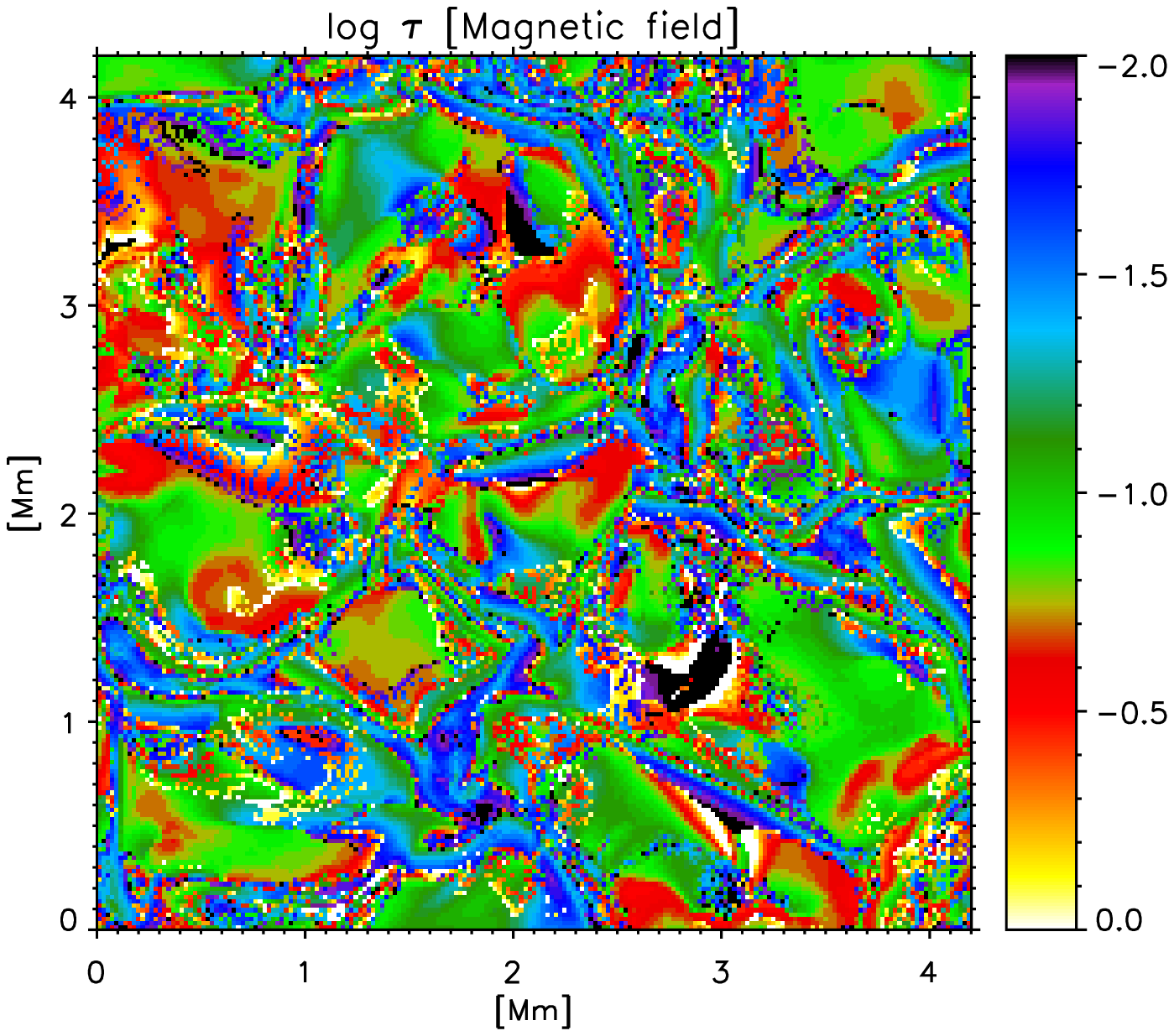}}
\resizebox{0.42\hsize}{!}{\includegraphics{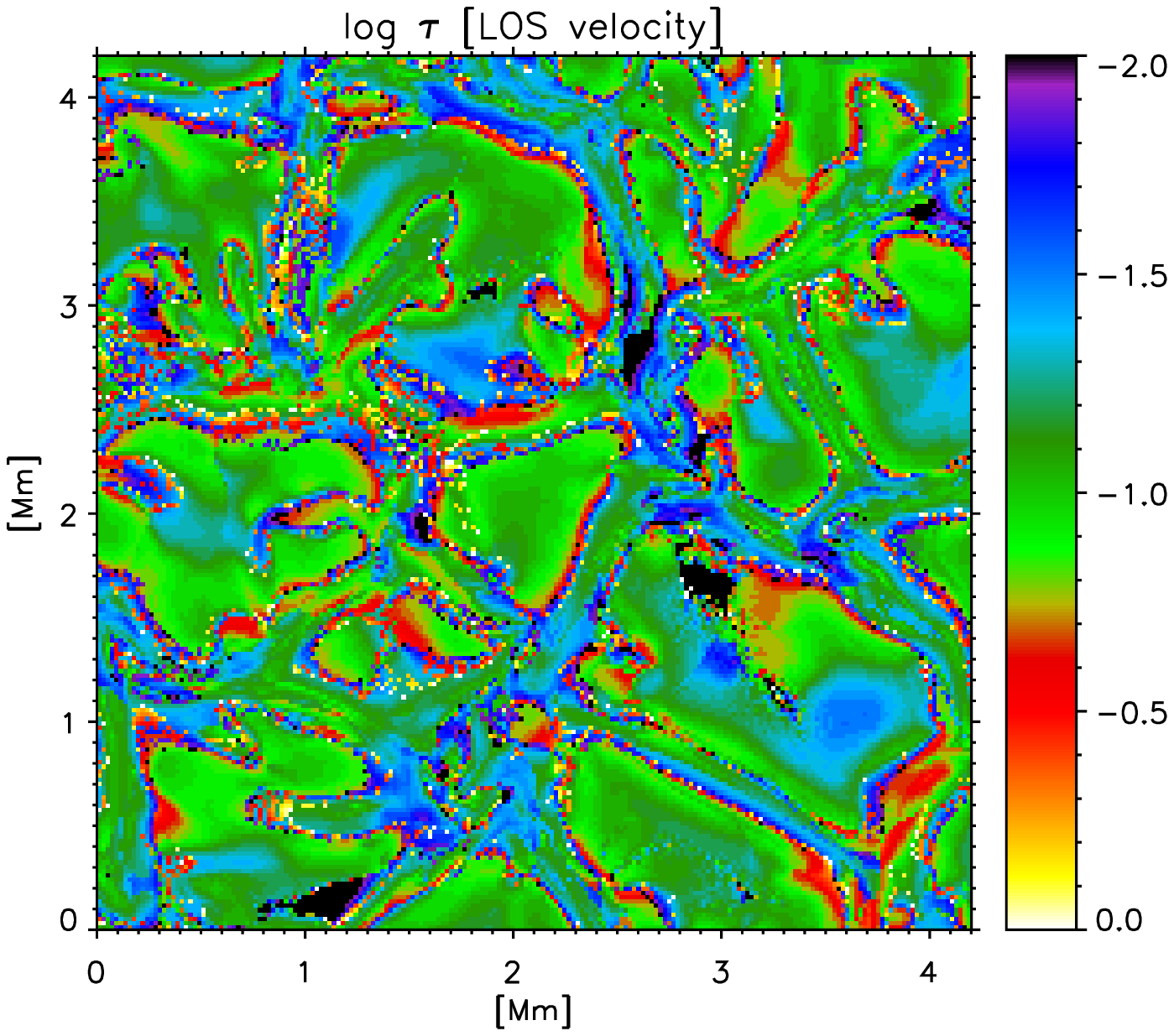}}
\resizebox{0.42\hsize}{!}{\includegraphics{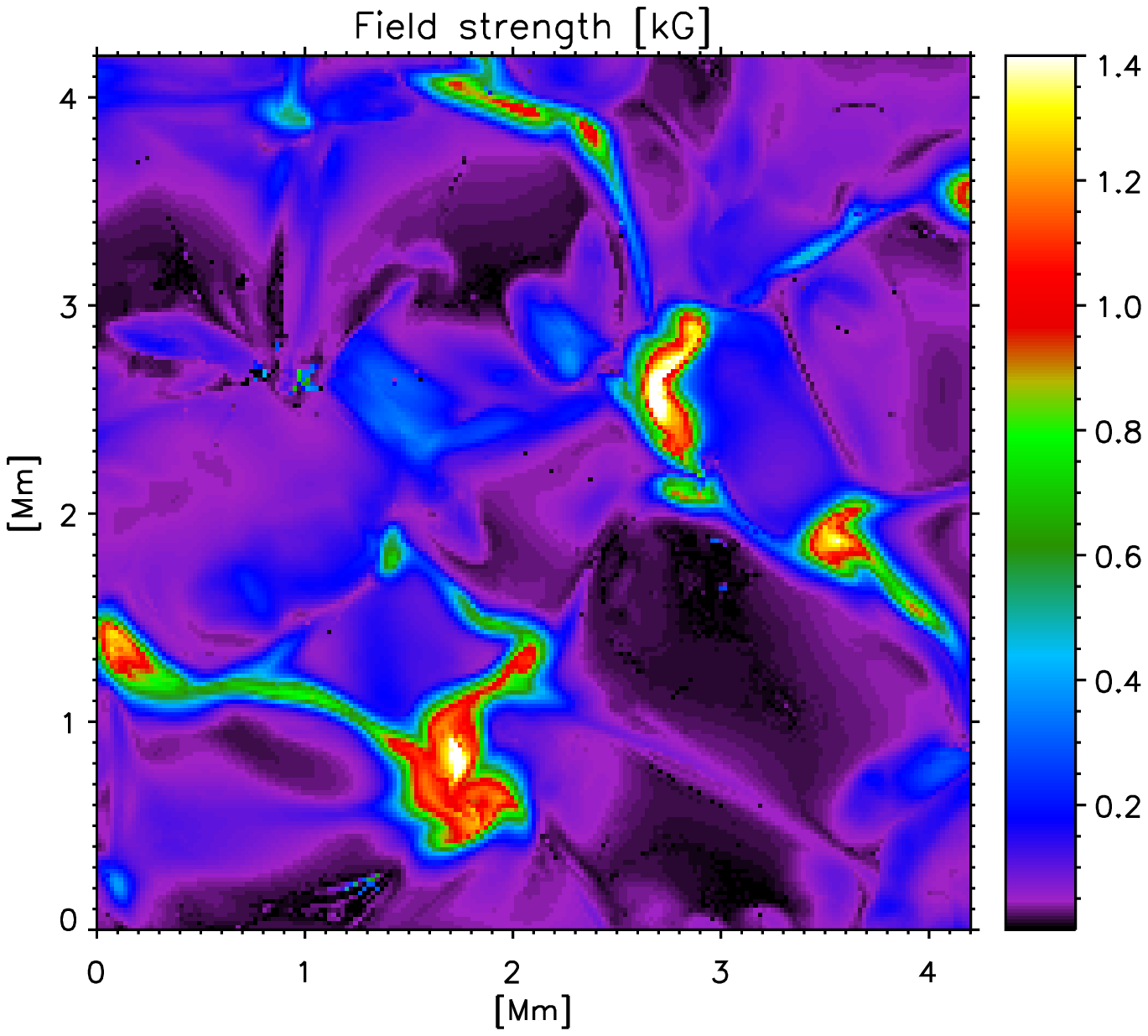}}
\resizebox{0.42\hsize}{!}{\includegraphics{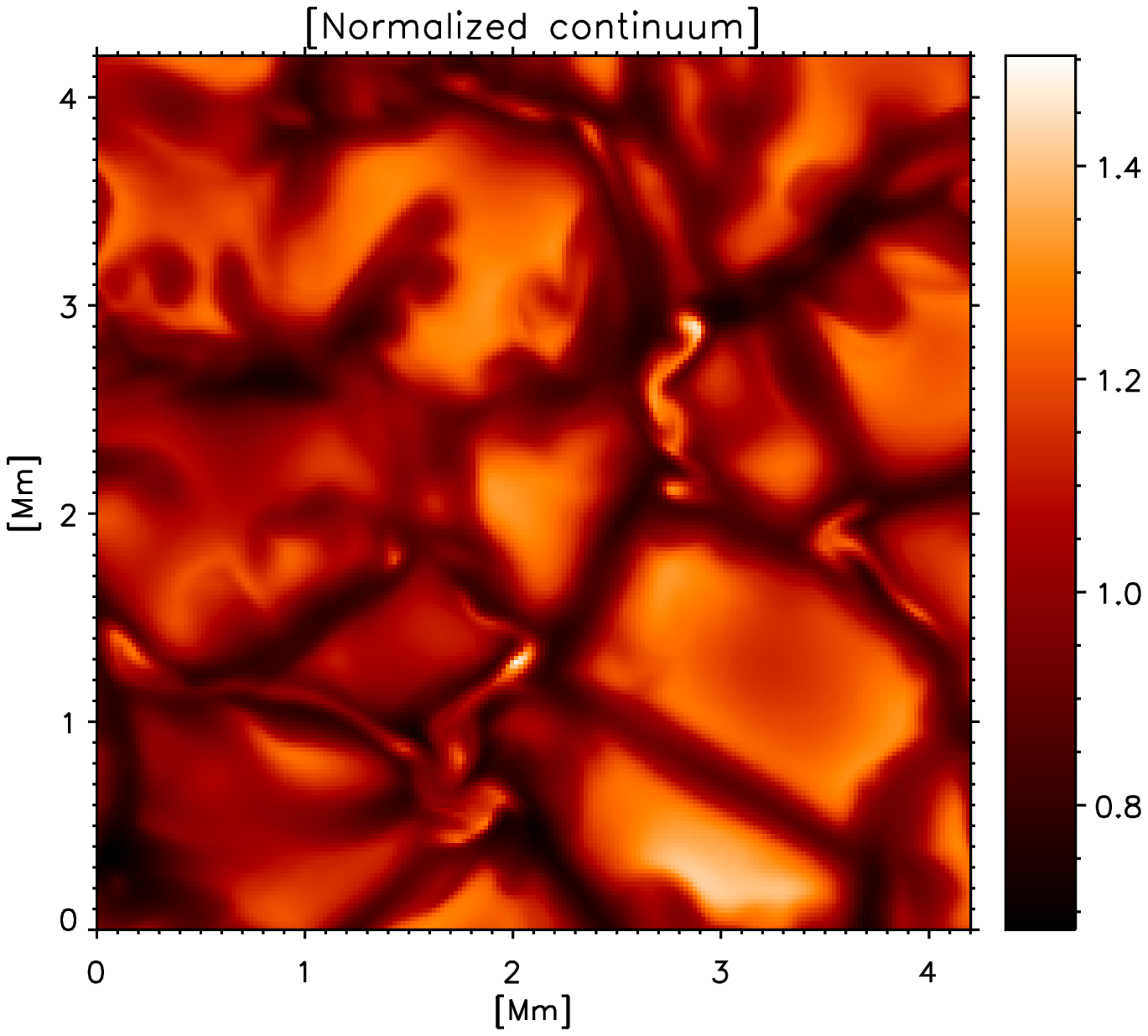}} 
  \caption{{\em Top:} Optical depths at which the inferred ME
  parameters coincide with the real magnetic field strength and LOS
  velocity stratifications (left and right, respectively). {\em
  Bottom:} Magnetic field strengths retrieved from the ME inversion
  and normalized continuum intensities (left and right,
  respectively).}
\label{Fig6} 
\end{figure*}

Figure \ref{Fig5} demonstrates that the ME model parameters coincide
with the real stratifications at specific optical depths. This allows
us to define the effective ``height of formation'' of the ME parameters.
 
Formation-height maps have been calculated by taking the optical depth
at which the stratification is closer to the inferred ME parameter.
The formation heights are constrained to be in the range from
$\log\tau=0$ to $-2$ because this interval includes most of the layers
to which the \ion{Fe}{i} lines are sensitive. When more than one value
of the MHD stratification coincides with the corresponding ME
parameter, we select the one located deeper in the atmosphere. The
optical depth of the minimum (or maximum) of the MHD stratification is
taken if the ME parameter is smaller (or larger) than all
stratification values. Even in that case, the formation height is not
allowed to go outside of the interval from $\log\tau=0$ to $-2$.

Figure \ref{Fig6} shows the results for the magnetic field strength
and the LOS velocity. For convenience, we also display the continuum
image and the field strengths retrieved from the inversion. Different
colors indicate different atmospheric layers. There are clear
differences between the two maps: the granular centers are
predominantly green ($\log\tau\in[-0.7,-1.2]$) in the velocity map and
green-red ($\log\tau\in[-0.3,-0.7]$) in the field strength map,
demonstrating that the ME inversion extracts the velocities from
higher optical depths than the magnetic fields, at least in
granules. The intergranular lanes tend to show blue colors in the two
maps ($\log\tau\in[-1.3,-1.7]$). The sharp red-yellow discontinuities
observed at the border of granules in the velocity map have little
meaning since they are a result of the way we estimate formation
heights. These regions have zero velocities (Fig.~\ref{Fig1}) and also
zero velocity gradients along the LOS, which makes our algorithm to
select layers close to the bottom of the photosphere (limited to $\log
\tau = 0$).  Both maps exhibit significant pixel-to-pixel
differences, especially the field strength map. This ``noise''
is due to MHD stratifications with many jumps in the vertical
direction.

In summary, ME inversions provide results that cannot be
assigned to a constant optical depth. The same parameter may show
formation-height differences of up to $1-1.5$ dex across the FOV. 
Also, the heights to which the ME parameters refer change depending on
the parameter, as predicted by \cite{1996SoPh..164..169D} and
\cite{almeida}. In the case of the \ion{Fe}{i} 630.2~nm 
lines, we find mean optical depths of $\log\tau = -1.0$ and 
$-1.1$ for the LOS velocity and the field strength, respectively. 
This includes granular and intergranular regions. If only 
intergranular regions are considered, the mean optical depths
move $\sim$~0.2~dex toward higher layers. The rms variation of the
formation heights is 0.4 and 0.5 dex, respectively.

\section{Inversion results}
\label{results2}

In what follows we compare the ME inversion results with the original
MHD models. To that end we use the atmospheric parameters of the
simulations at $\log\tau=-1$. This layer corresponds to the average
formation height of the ME parameters. As such, it represents the best
choice for a ``reference model''.

\begin{figure}[t] 
\centering 
\resizebox{0.98\hsize}{!}{\includegraphics{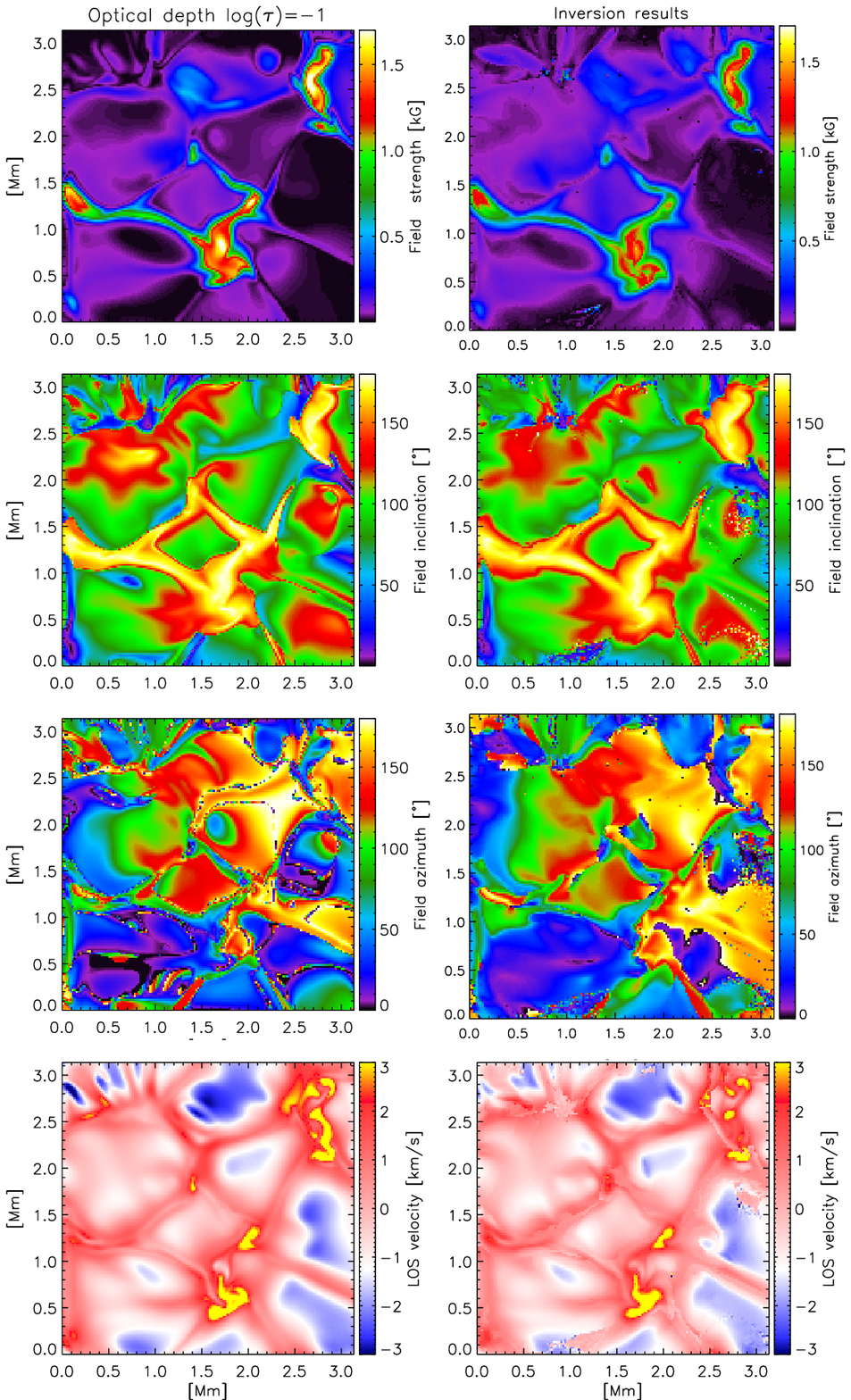}} 
  \caption{From top to bottom: magnetic field strength, inclination,
  azimuth, and LOS velocity. The left column represents the MHD
  parameters at $\log\tau=-1$. The right column shows the results of
  the ME inversion of the \ion{Fe}{i} lines at 630.2~nm.}
\label{Fig7} 
\end{figure}

Figure \ref{Fig7} displays maps of the magnetic field strength,
inclination, azimuth, and LOS velocity in the reference model (left
column) and the models retrieved from the inversion (right column).
To better visualize the details we only show a small area of about
9~Mm$^2$. The strong resemblance between the reference parameters and
the ME models is obvious: the shapes of the different structures are
well reproduced and only small differences can be
recognized. Sometimes the inversion yields bad results for the
inclination and azimuth, but this happens mainly in areas with weak
polarization signals.

\begin{figure}[t] \centering % 
  \resizebox{\hsize}{!}{\includegraphics{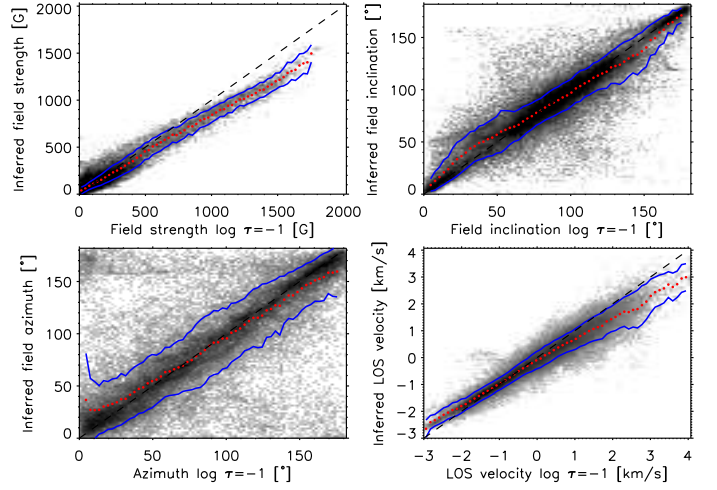}}
\caption{Scatter plots of the magnetic field strength, inclination,
  azimuth and LOS velocity inferred from the ME inversion vs the MHD
  parameters at $\log\tau=-1$. The dashed lines represent
  one-to-one correspondences. The red dots are the mean values of the
  parameters in small, evenly-spaced bins along the X-axis.  The blue
  lines represent the rms fluctuations of the parameters in those
  bins.}  \label{Fig8} \end{figure}

From a visual inspection of the maps, one can say that the 
ME inversion is able to determine the magnetic field vector
satisfactorily. Even structures with field strengths as low
as 100~G are well recovered. To make more precise statements,
Fig.~\ref{Fig8} shows the parameters inferred from the fit vs the 
MHD parameters at $\log\tau=-1$. These scatter plots allow us to
estimate the uncertainties that can be expected from the use of the ME
approximation, since no noise has been added to the profiles
(Sect.~\ref{inversion}).

As can be seen, the scatter is larger for the magnetic field
inclination and azimuth. The mean values of the
parameters\footnote{The average values have been calculated by taking
bins along the X-axis of size 28~G, 3\degree, and 115~\ms, depending
on the physical quantity.}  (red dots) show that the magnetic field
strength is really close to that in the reference model from 0 to
500~G. For stronger fields the retrieved values are slightly
underestimated, although the deviation does not exceed $\sim$250~G.
The red lines indicate maximum rms fluctuations of about $\sim$80~G
in the whole range of strengths. The inclination shows rms
fluctuations smaller than 10\degree\/ for vertical fields and
15\degree\/ for inclined fields. The rms variation in the azimuth 
is about 15\degree\/. The LOS velocity panel shows that the 
retrieved velocity is some 200-300~\ms\/ smaller than the 
reference velocities for receding flows (intergranular lanes). 
The rms values are smaller than $\sim$~500~\ms\/ in the 
full velocity range.

The scatter observed in the various panels of Fig.~\ref{Fig8}
originates from the use of ME model atmospheres (unable to fit
asymmetric Stokes profiles) and the pixel-to-pixel variations of the
formation height of the ME parameters, as explained in the previous
section. The deviation of the ME field strengths from a one-to-one
correspondence with the MHD models can easily be understood by looking
at the top panel of Fig.~\ref{Fig6} and recalling that we have chosen
the atmospheric layer at $\log\tau=-1$ as a reference. In those
spatial locations where the optical depth assigned to the retrieved ME
parameter is smaller than the optical depth of the reference layer,
the resulting field strength will ``apparently'' be
underestimated. These spatial locations are associated with strong
flux concentrations; in the MHD models they spread out with height,
therefore we retrieve weaker fields.

The rms differences between the inversion results and the MHD models
tell us how much a single ME parameter could deviate from the real
value, but only if the {\em mean differences} are zero or close to
zero. Non-zero mean differences indicate that {\em systematic} errors
exist. For this reson, when the mean difference is larger than the
standard deviation, the former should be preferred as a better
estimate of the true error.

\begin{figure}[!t] \centering %  
 \resizebox{\hsize}{!}{\includegraphics{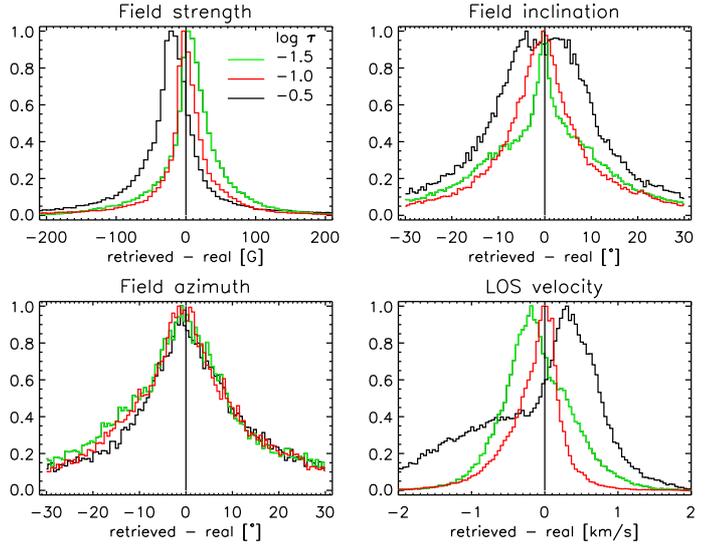}}
 \caption{Normalized histograms of the differences between the
inferred ME model parameters and the real ones taken at different optical
depths.} \label{Fig9} \end{figure}

The choice of $\log \tau = -1 $ for the reference model is appropriate
because it produces the smaller mean and rms values on average. To
illustrate this, Fig.~\ref{Fig9} represents histograms of the
differences between the inferred parameters and the MHD model at three
optical depths ($\log\tau=-0.5,-1,-1.5$, coded in black, red, and
green, respectively).

For the magnetic field strength, the histogram corresponding to
$\log\tau=-1$ peaks around zero. The maximum shifts toward negative
values when the inversion results are compared with deeper layers
(fields are underestimated on average) and toward positive values
when the comparison is made with higher layers (over-estimating the
strength). The full width at half maximum (FWHM) of the distribution
is about 30~G for $\log\tau=-1$, and increases up to $\sim$45 and
$\sim$50~G for $\log\tau=-1.5$ and $-0.5$, respectively. These effects
are less pronounced for the field inclination: the peaks of the
histograms are located at zero and the FWHM varies from 6\degree\/
($\log\tau=-1.5$) to $\sim$13\degree\/ and $\sim$23\degree\/
($\log\tau=-1$ and $-0.5$, respectively). The larger FWHMs originate
from the extended wings of the distributions.  The azimuth histogram
does not vary when the comparison is made with different optical
depths. In this case, the FWHM is about 20\degree.

The histograms of the LOS velocity differences show larger variations.
The one corresponding to $\log\tau=-1$ has the smaller FWHM
($\sim500$~\ms). It also features a long tail toward negative values
caused by pixels located in intergranular lanes.  The asymmetry of the
histograms around the location of the peaks changes dramatically when
we compare the inversion results with different atmospheric
layers. For instance, if the reference layer is taken at
$\log\tau=-0.5$, the histogram is a clear combination of two different
distributions, one representing granular centers (higher and narrower)
and the other representing intergranular lanes (smaller in amplitude 
and broader).

\section{Summary and conclusions}
\label{conclus}

In this paper we have analyzed radiative MHD simulations of the quiet
Sun. We have used them to synthesize Stokes line profiles in three
different spectral regions (525.0, 617.3, and 630.2~nm). The
comparison of the synthetic profiles with the FTS atlas suggests that
the simulations describe quite satisfactorily the physical conditions
of the solar photosphere, although the MHD models are slightly hotter
than the HSRA around $\tau=1$.

After synthesizing the Stokes profiles, the applicability of ME
inversions to high spatial resolution observations has been
examined. We have considered the case of the \ion{Fe}{i} lines at
630.2~nm. The analysis of the profiles by means of ME inversions has
allowed us to characterize the uncertainties that can be expected from
the ME approximation. For this reason, the synthetic profiles were
not degraded by noise, instrumental effects, or spatial resolution.
 
The main limitation of ME inversions is that they provide constant
atmospheric parameters, whereas the MHD models feature physical
properties that change with height. This limitation means that ME
models are unable to reproduce spectral line
asymmetries. Consequently, the ME inferences cannot be assigned to a
specific optical depth. Depending on the conditions of the atmosphere,
the retrieved ME parameters sample different layers.

However, from a statistical point of view we conclude that ME
inversions provide fair estimates of the physical conditions
prevailing at $\log\tau \sim -1$. The rms uncertainty is smaller
than 30~G for the magnetic field strength, 13\degree\/ and 20\degree\/
for the field inclination and azimuth, and 500~\ms\/ for the LOS
velocity. Thus, ME inversions are appropriate for statistical analyses
of the solar photosphere. This being said, it is important to realize
that the errors may be large for individual pixels, even if the
best-fit profiles reproduce the observations satisfactorily (the field
strength in case {\em a} of Fig.~\ref{Fig5} is a good example of this).

 Finally, we want to stress that the uncertainties associated with 
the ME approximation are larger than those due to photon noise
\citep{2006ASPC..358..197O,2010ApJ...711..312D}. However, the noise has
another undesirable effect: it hides the weaker polarization 
signals. This fact has not been considered in our study.

%________________________________________________________________
\begin{figure*}[!t] \centering %  
  \resizebox{0.98\hsize}{!}{\includegraphics{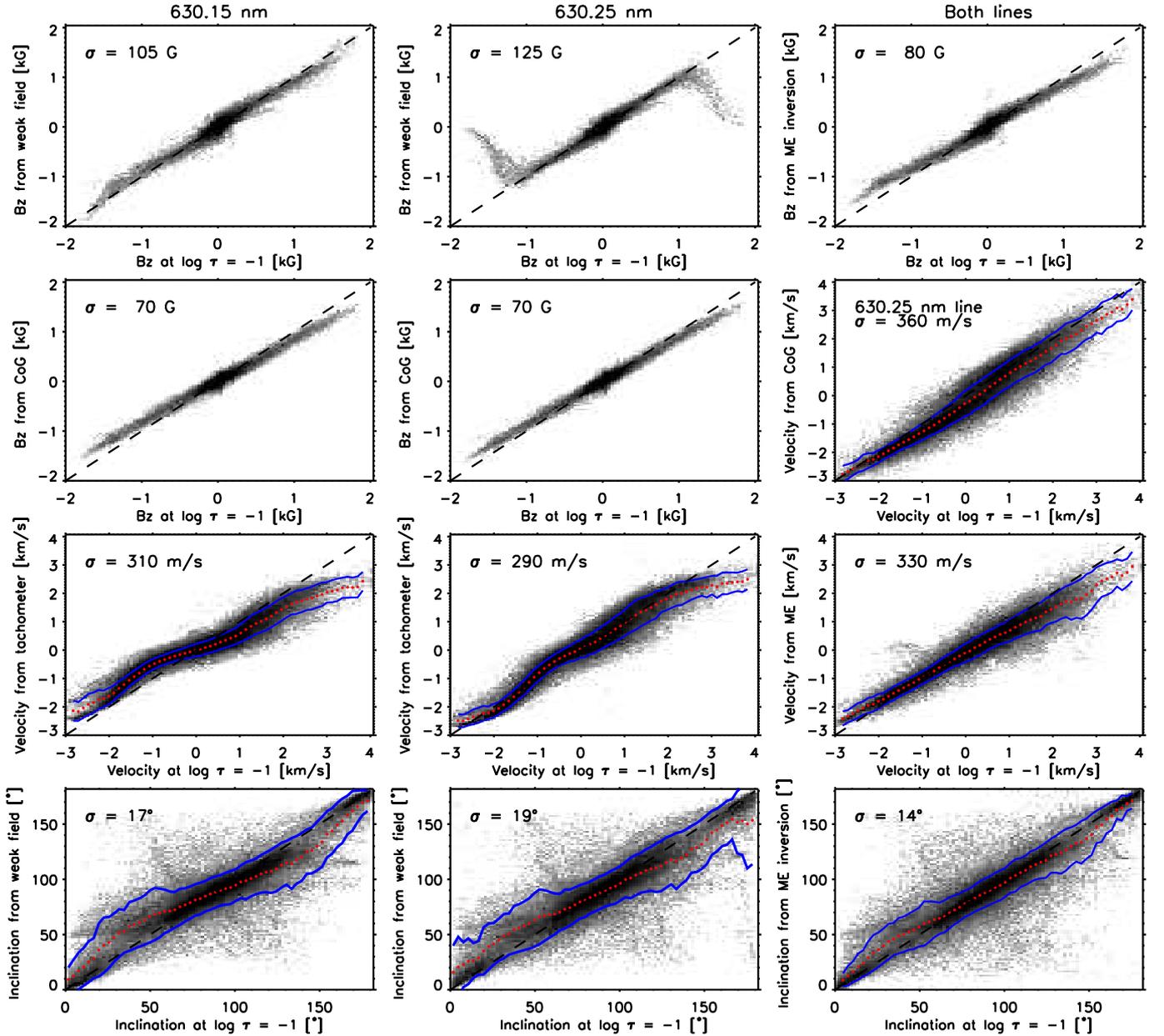}}
  \caption{Scatter plots of the longitudinal magnetic field, $B_{z}$
  (first two rows), of the LOS velocity (third row), and
  of the magnetic field inclination (fourth row), as functions of their
  corresponding values in the MHD simulations at $\log\tau = -1$. The
  first column refers to results obtained with the 630.15 nm line; the
  second column displays results obtained with the 630.25 nm line;
  results with the two lines at the same time are shown in the third
  column. Color and line codes are the same as in Fig.\ \ref{Fig8}. The mean value of 
the rms difference between the inferences and the simulations at $\log\tau = -1$ 
is given in the upper left corner of each panel.} 
  \label{FigApp}
\end{figure*}

\appendix 

\section{Milne-Eddington vs classical proxies}
\label{proxies}

The study of the solar atmosphere relies on the availability of
precise magnetic fields and LOS velocities. Thus, one needs robust
diagnostics in order to extract this information from the Stokes
spectra. Classical methods such as tachogram techniques, the weak
field approximation, and the center-of-gravity technique represent an
alternative to Stokes inversions.

For some of these methods, the {\em random} errors induced by photon
noise have been estimated not to exceed $\sim$~20~m~s$^{-1}$ in the
case of the LOS velocity or $\sim$~10~G in the case of the magnetic
flux (see e.g., Scherrer et al.\ 1995; Scherrer \& SDO/HMI Team 2002;
Mart\'inez Pillet 2007). However, like in the case of ME inversions,
{\em systematic} uncertainties coming from the hypotheses underlying
the technique are expected to be larger than the random errors
themselves. A thorough study, similar to the one we have performed for
the ME technique, is thus in order. We carry out such an analysis in
this Appendix for the Fourier tachometer technique
\citep{1978fsoo.conf..189B,1981siwn.conf..150B}\footnote{We 
in fact use the formula proposed by \cite{1992PhDT.........3F}.}, the
center-of-gravity method \citep{1967AnAp...30..513S,1979A&A....74....1R}, 
and the weak field
approximation \citep{1992soti.book...71L,2004ASSL..307.....L}. 

Both the center-of-gravity method and the weak field approximation
are applied to the whole profiles while the Fourier tachometer uses
only four wavelength samples across the Stokes $I$ profile ($-9$,
$-3$, $3$, and $9$ pm). The center-of-gravity technique extracts the
longitudinal component of the magnetic field, $B_{\rm z}$, from the
separation between the barycenters of the Stokes $I+V$ and $I-V$
profiles. $B_{\rm z}$ can also be obtained with the weak field
approximation through a proportionality between the Stokes $V$ profile
and the wavelength derivative of Stokes $I$. The transverse component
of the field, in turn, is derived through a proportionality between
Stokes $L$ and the second wavelength derivative of Stokes $I$.
\footnote{Stokes $L$ is the total linear polarization, 
$L=\sqrt{Q^2+U^2}$.} Regression fits are used between the circular
(linear) polarization profiles and the first (second) derivatives of
the intensity profiles for increased accuracy. Then, the magnetic
inclination is obtained from the ratio between the transverse and
longitudinal components of the field.

Figure \ref{FigApp} summarizes the results. Each column refer to a
different set of lines: \ion{Fe}{i} 630.15~nm (left), \ion{Fe}{i}
630.25~nm (middle), and the two lines simultaneously considered (i.e.,
ME inversion; right). The labels on the ordinates are
self-explanatory, while the abscissae give the values of the
corresponding quantities at $\log\tau =-1$. 

The less accurate method turns our to be the weak field approximation:
the inferred magnetic inclinations show larger rms fluctuations than
the ME ones, and the longitudinal component of the field displays a
clear saturation for fields stronger than 1000 -- 1100 G when
calculated with the line at 630.25 nm. For $B_{\rm z}$ values above 1
kG, the weak field inferences resulting from the 630.15 nm line seem
to be closer to the MHD values at $\log \tau = -1$ than the ME
ones. Nevertheless, they present a larger scatter. The
center-of-gravity method looks very robust and, indeed, it presents
less scatter than ME inversions in the case of the longitudinal field
component and the LOS velocity (only the results from the 630.15 nm
line are shown; the results for 630.15 nm are very similar). The good
performance of the center-of-gravity method was noticed earlier by
\citet{1993SoPh..146..207C} and \citet{2003ApJ...592.1225U}.
Unfortunately, this technique does not provide information about the
field inclination. The LOS velocities resulting from the tachometer
are fairly comparable to those of the ME inversion. The scatter is
similar in both cases. 

In summary, the ME inversion seems to be the more complete and
accurate technique although none of the others can be discarded. In
particular, a combination of the center-of-gravity technique for
calculating $B_{z} $ and $v_{\rm LOS}$ along with the weak field
approximation for the magnetic inclination may represent a suitable
alternative which is much less expensive in terms of computing
resources. It is important to note, however, that the results of our
study are only valid when the magnetic field is spatially resolved. 
Further investigation is needed to check the applicability of these 
techniques when the field is unresolved. This additional investigation 
is important in view of the theoretical deviations predicted by 
\cite{2004ASSL..307.....L}.

\begin{acknowledgements}
This work has been funded by the Spanish MICINN through 
projects AYA2009-14105-C06-06 (in\-clu\-ding European FEDER 
funds) and PCI2006-A7-0624, by Junta de Andaluc\'{\i}a through 
project P07-TEP-2687, and by the Japan Society for the Promotion 
of Science.
\end{acknowledgements}

\end{document}